\documentclass[prl,aps,showpacs,twocolumn,floats]{revtex4} \input epsf
\usepackage{times}
\usepackage{graphicx,color} 
\usepackage{amsmath,amssymb}
\usepackage{bm}

\begin{document}

\title{Stretching force dependent transitions in single stranded DNA}
\author{Kulveer Singh$^{1}$}
\author{Surya K. Ghosh$^{1}$}
\author{Sanjay Kumar$^{2}$}
\author{Anirban Sain$^{1}$}
\email{asain@phy.iitb.ac.in}
\affiliation{$^1$ Department of Physics, Indian Institute of Technology, Bombay,
  Powai, Mumbai-400 076, India.\\
$^2$ Department of Physics, Banaras Hindu University, Varanasi 221 005, India}

\date{\today}

\begin{abstract}
Mechanical properties of DNA, in particular their stretch dependent extension and their
loop formation characteristics, have been recognized as an effective probe for understanding 
the possible biochemical role played by them in a living cell. Single stranded DNA 
(ssDNA), which, till recently was presumed to be an simple flexible polymer continues 
to spring surprises. Synthetic ssDNA, like polydA (polydeoxyadenosines) has revealed an 
intriguing force-extension (FX) behavior exhibiting two plateaus, absent in polydT 
(polydeoxythymidines) for example. Loop closing time in polydA had also been found to 
scale exponentially with inverse temperature, unexpected from generic models of 
homopolymers. Here we present a new model for polydA which incorporates both a 
helix-coil transition and a over-stretching transition, accounting for the two 
plateaus. Using transfer matrix calculation and Monte-Carlo simulation we show that 
the model reproduces different sets of experimental observations, quantitatively. 
It also predicts interesting reentrant behavior in the temperature-extension 
characteristics of polydA, which is yet to be verified experimentally.
\end{abstract}

\maketitle
In order to understand how the mechanical properties of DNA and RNA influence 
biological processes like transcription and translation in living cells and viruses, 
these biopolymers are stretched in vitro to study their nonlinear elasticity 
and their internal structure. ssDNA, despite receiving less attention compared to 
double stranded DNA (dsDNA) \cite{smith92,bustamante94}, has recently attracted a 
lot of interest. Smith et al. \cite{smith96} showed that the FX diagram
for wild type $\lambda$-phase ssDNA can be described by FJC model only at low force 
($<10pN$). To explain the behavior at higher force they used a {\em modified} freely 
jointed chain (mFJC) model \cite{smith92,smith96} with stretch dependent kuhn length.
Subsequently, it was discovered that synthetic ssDNA have interesting sequence dependent 
properties. For example, polydA was found to have a higher bending rigidity than 
polydT and as a consequence polydA takes relatively longer time to form a loop 
\cite{goddard2000prl}. Further, the loop formation time varies exponentially 
with inverse temperature which can neither be explained by FJC nor worm like 
chain (WLC) model, which are generic models for flexible and semiflexible 
polymers, respectively. 
This behavior was attributed to strong stacking interaction among pyrimidine
bases in polydA \cite{goddard2000prl,goddard02BiophysJ,chen04}. Subsequently, 
FX characteristics of polydA revealed that it undergoes two successive 
transitions under external stretching \cite{ke,prl2010} generating two distinct 
plateaus in the force extension (FX) curve. The first one at $\sim 23pN$ force 
was proposed to be a helix to coil transition in which the inter-base stackings 
are broken and helical polydA segments transform to a polydA coil. The second 
transition at $\sim 114pN$ force was attributed to the over-stretching of the 
constituent bases. It was conjectured \cite{smith96,ke} that over-stretching 
results from the conformational change of the sugar molecules from C3'-{\em endo} 
to C2'-{\em endo} pucker conformation.

Zimm-Bragg model \cite{Zimm-Bragg} which was originally proposed to explain 
temperature driven helix-coil transition in proteins, has been used to study 
the force driven helix-coil transition in polydA. Theoretical models 
have used mean field  approximation \cite{buhot} and exact evaluation of 
partition function \cite{pincus} to explain the first plateau experimentally
seen \cite{seol07} in the FX diagram of polydA at low forces 
($< 60pN$). But the second plateau, involving the over-stretching transition 
is beyond the scope of these models. 
Overstretched in dsDNA has been studied using Ising like two state models by 
various groups \cite{amir,cluzel}. Here we propose a model which quantitatively 
reproduce both the force driven behavior as well as zero force conformational 
fluctuations like the loop formation time, as observed experimentally. Double 
plateau behavior has also been addressed before \cite{sanjay,sanjaySoftRev}, 
theoretically, albeit using a lattice model. But lattice models generically 
underestimate entropic effects and also quantitative comparisons to experiments 
were not possible. 

{\em Model:} We model polydA as a chain of connected segments, each representing a 
nucleotide. Length of a segment (bond) represent the phosphate to phosphate distance 
in the ssDNA backbone. The bond length can be $l_h,l_c$ or $l_s$ depending on whether 
the segment is in the helix ($l_h=0.37nm$ \cite{seol07}), coil ($l_c=0.59nm$ 
\cite{bustamante94}), or overstretched-coil state ($l_s=0.7nm$ \cite{ke}). While the 
value of $l_c$ has been most well documented \cite{bustamante94} in the literature,
$l_h$ is determined \cite{seol07} by noticing the $1.6$ times increase in contour 
length of polydA upon helix to coil transition. Here $l_h$ of course is the projected 
length of the helical contour on the central axis of the helix. Further, $l_s=0.7nm$ 
has been inferred from the maximum extension reached by the polydA chain under very 
high force ($\sim 600pN$) in Ref\cite{ke,prl2010}. Incidentally, $l_s=0.7nm$ also 
matches with the distance between the consecutive phosphates when the deoxyribofuranose 
ring is in the C2'-{\em endo} pucker conformation. In our model, state of a segment 
is characterized by $(\mu, S)$, where $ \mu $ can take values $0$ or $1$, corresponding
to coil or helix state, respectively. Once in the coil state $\mu=0$, there are 
two possible states: the normal coil or the overstretched coil, corresponding to 
bonds lengths $l_c$ and $l_s$. These two states are represented by $S=-1$ and $S=+1$,
respectively. Hence, there are only three possible states $(1,-1), (0,-1)$ and 
$(0,1)$, corresponding to helix, coil and overstretched-coil.  The Hamiltonian of 
this three state model of polydA (omitting the external force) is
\begin{eqnarray}
H_0&&= \sum_{i=1}^{N} \Big( 2(1-\mu_{i})\mu_{i+1}\Delta w+\mu_{i+1}\Delta f  \nonumber \\ 
 &&-J S_{i} S_{i+1} (1-\mu_{i})(1-\mu_{i+1})+h S_{i}(1-\mu_{i}) \Big) 
\label{HamiltonianwithoutF}
\end{eqnarray}
This Hamiltonian incorporates two transitions, actually cross-overs, since the model
is one dimensional.  First part of the Hamiltonian involving $\Delta w$ and $\Delta f$, 
is the simpler version of the original Zimm-Bragg model \cite{Zimm-Bragg}, forwarded by 
the authors themselves and later used by Tamashiro et al. \cite{pincus} in the context
of forced DNA. In this model the necessary hydrogen bonding, that is required 
for the formation of helical domain, takes place between adjacent bases (segments), instead 
of the $i-th$ and $(i+4)-th$ bases as in the original Zimm-Bragg model \cite{Zimm-Bragg}. 
The weight of the configurations $cc,\; hc,\; ch,\; hh$, in this model, are given 
by $1, 1, \sigma s$ and $s$ respectively. Here $ \sigma = e^{-2 \Delta w/k_B T}$ and
$ s=e^{-\Delta f/k_B T}$, where $2 \Delta w $ is the interfacial energy between the 
helical and the coil domains and $ \Delta f$ is the difference of energy between the helix 
and the coil state. The asymmetry between $ch$ and $hc$ arises because a segment (its 
oxygen atom) can engage either its right neighbor (its hydrogen atom) or its left neighbor 
to make a hydrogen bond (for details see Ref\cite {Zimm-Bragg}).
Second part of the Hamiltonian is an Ising Hamiltonian, which will be invoked, when the
segments are in the coiled or overstretched coil state i.e., when $\mu=0$. $J$ measures 
the correlation energy between an adjacent coil and overstretched coil states. $2h$ is 
the energy difference between a coil and a overstretched coil state. The parameters
$\sigma$ and $J$ are often called the cooperativity parameters of the respective transitions.

We considered the ssDNA chain to be semi-flexible, in which the helical domains have very 
large persistence length, where as the coiled and overstretched-coil domains have 
small persistence length. Discretized worm-like-chain hamiltonian for the system is
\begin{eqnarray}
&& \beta H_{bend} =\sum_{i=1}^{N-1}\Big\{\frac{a_h}{2}\mu_{i}\mu_{i+1}
+ (1-\mu_{i})(1-\mu_{i+1})\times \nonumber \\ 
&& [\frac{a_c}{8}(1-S_{i})(1-S_{i+1})+\frac{a_s}{8}(1+S_{i})(1+S_{i+1})]\Big\}\times 
\nonumber \\
&&(1-\cos\theta_{i,i+1}). 
\end{eqnarray}
Here, $\theta_{i,i+1}$ is the angle between the bond vectors $\vec t_i$ and 
$\vec t_{i+1}$, where  $\vec t_i={\bf R}_{i+1}-{\bf R}_{i}$ and $ {\bf R}_{i}$ 
is position vector of $ i-th $ monomer. The hamiltonian has been constructed
in a way such that different bending rigidities are associated when neighboring
bonds are of identical type, i.e., $hh,cc$ or $ss$. It amounts to assuming that
helical, coil and stretched-coil domains of the polymer behave like worm-like-chain.
$a_h,a_c$ and $a_s$ are the respective persistence lengths. We choose $a_h=12nm, 
\;a_c=1.5nm$ and $a_s=1.5nm$, i.e., a relatively large persistence length for 
the helix (still much smaller than that of $dsDNA$, about $~50nm$). Theoretical 
models \cite{pincus,seol07} assume $a_h$ to be infinite. Our assumption 
that even ssDNA coils have a small persistence length is in agreement with Seol 
et al. 's \cite{seol04} FX data on polyU where they obtained $a_c\sim 1nm$. 
Also Smith et al \cite{smith96} reported  $a_c\sim 1.5nm$ for wild type 
$\lambda$-phase ssDNA, using mFJC model.

 For calculation purpose, 
we substitute inter-bond angles in terms of bond vectors:
$\frac{a}{2}(1-\cos\theta_{i,i+1})=\frac{a}{4}(\hat t_{i+1}-\hat t_i)^2$.
Finally, including external stretch ${\bf F}$, the total hamiltonian is 
\begin{eqnarray}
H_{total} &=& H_0+H_{bend} -{\bf F}.({\bf R}_{N}-{\bf R}_{0})
\label{HamiltonianwithF}
\end{eqnarray}
The force dependent term above can be expressed as ${\bf F}.\sum_{i=1}^N 
l_i \hat t_i$ in terms of the bond vectors and their respective lengths $l_i$,
where $l_i$ could be $l_h,l_c$ or $l_s$ depending on the internal state 
$(\mu,S)$ of the bond.
Finally, the partition function is 
$Z(F)=\int \Pi_{i=1}^N d\hat{t}_i \sum_{\{\mu_i,S_i\}}
\exp\left[-\beta (H_0+H_{bend} - F R_x) \right]$,
where $R_x=({\bf R}_{N}-{\bf R}_{0}).\hat x$ is projection of the end-to-end 
distance along the force ${\bf F}=F\hat x$. Using transfer matrix technique we 
can write $Z(F)$ for the forced chain as,
\begin{eqnarray}
Z=\sum_{\mu_1,S_1,\hat{t}_1;\mu_N,S_N,\hat{t}_N}\langle{\mu_1,S_1,\hat{t}_1}|T^{N-1}|{\mu_N,S_N,\hat{t}_N}\rangle
\label{Eqz}
\end{eqnarray}
Here the transfer matrix $T$ is an $3mn \times 3mn$~matrix where the internal state space
is $3$ dimensional, corresponding to the states $(1,-1),(0,-1),(0,1)$ and the orientation 
space $\hat{t}_i(\phi,\theta)$ has been discretized into $m\times n$ bins. 
In Eq.\ref{Eqz}, $Z(F)$ is obtained as a weighted sum over all the 
matrix elements of $T^{N-1}$. 
We cannot exploit the simplifications normally arising from periodic 
boundary condition because, here, the polymer 
has one of its ends fixed and the other end stretched by a force.  
The details of this calculation can be found in Ref\cite{curvedpre}. 
After calculating $Z(F)$ we can compute $\langle R_x \rangle$ as a function of 
force using $\langle R_x \rangle = \frac{1}{\beta} \frac{\partial ln Z}{\partial F}$.
This is shown in Fig.1.

\begin{figure}[htb]
\includegraphics[width=7.5cm]{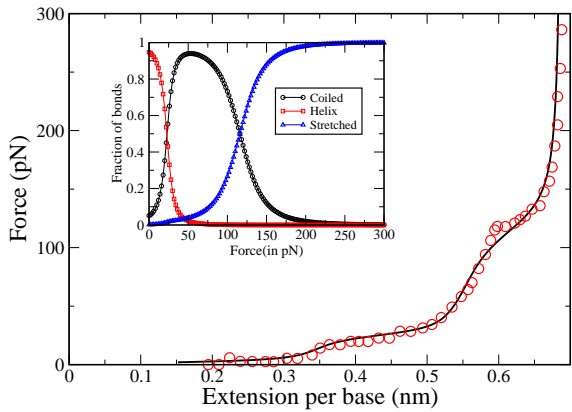}
\vspace{.25cm}
\caption{(color online) Force-Extension curve obtained from our model (solid line), using transfer matrix 
method, compared with the experimental result of Chen et al. \cite{prl2010} (circle). Such 
a double plateau feature was first reported by Ke et al and their data is very close to 
that of Ref\cite{prl2010} shown here. The inset shows how fraction of the different species, 
namely, helix(circle), coil(square) and stretched coil (triangle) portions change with 
increasing force, obtained from Monte-Carlo simulation of our model.  The intersection points 
between the helix and coil fractions mark the position of the first plateau ($\sim 23pN$) and that between 
the coil and overstretched-coil fractions mark the second plateau  ($\sim 114pN$) in the FX-plot.
Parameters used for our calculation are $2\Delta w=6.3k_B T,\; \Delta f=-4.93k_B T,\; J=0.44k_B T,\; 
h=1.5k_B T$, where $T$ is the room temperature. We used two different chain lengths, 
$N=33$ and $65$, to check $N$ independence of this plot. }
\label{fig.FX}
\end{figure}

{\em Force-extension behavior} :
In general the stretching force tends to align the chain along the force ($\hat x$), at the 
cost of entropy. It would also favor the individual bonds to have their highest possible 
bond lengths i.e., $0.7 nm$ in order to maximize the $FR_x$ term. This is achieved only
at very high force when the other terms in the Hamiltonian give in to the force term. But
at low and intermediate forces the other terms compete. Although helical segments are 
favored over coil segments (due to $s$), at low force and at room temperature, entropy has 
substantial contribution. As a result all the segments are not aligned along the force. 
That is why the extension per base, at very low force is about $0.2nm$ and not $0.37nm$
(see Fig.1).  But bending of the helical domain is disfavored by its large persistence length. 
Therefore, at low force, the entropy induced bends help 
some helical segments to convert to coil, which have higher internal energy but lower 
bending energy compared to helical segments.    
As the force rises, the $FR_x$ term 
enforces alignment as well as stretching of the bonds, leading to stacking-unstacking 
transition which adds extra length (that was previously curled up in helix) to $R_x$.
The abrupt nature of the helix-coil transition results from high cooperativity 
($\sigma=e^{-2 \Delta w/k_B T}=0.0015$ here) i.e., the 
coupling term $\mu_i\mu_{i+1}$ in the Hamiltonian. The width of the plateau, i.e., how 
much length is released at the transition, depends on how many segments are converted 
from the helical to the coil state. This comes from the abruptness of the jump, shown 
in the inset. We compute the area under the FX curve: $\int F.dx\sim 6k_BT$. Given that
$2h=3k_BT$ is the free-energy change for the second transition, the helix-coil transition 
accounts for a free-energy change of about $3k_BT$. 

Given that polydT and polyU do not have any significant stacking interaction, they might 
be expected to follow the $\Delta f=\Delta w= 0$ limit of our model. But 
polydT,polyU also have qualitatively different overstretching behavior than polydA; they 
overstretch very slowly with force (unlike the transition in polydA). PolyU has been 
already explained by a modified WLC model \cite{seol04} and we checked that even polydT
can also be explained by the same model but with different parameters. 

{\em Loop formation kinetics} :
Now we discuss the zero force conformational fluctuations resulting from our model. 
We focus on the observations
made by Goddard et al \cite{goddard2000prl} on loop formation properties of short ($N=8-30$ bases) 
polydA and polydT chains. They had attached complementary base sequences TTGCC and AACGG at the two
ends of a polydA/polydT strand and attached flurophore ($F$) and quencher ($Q$) molecules at the ends.
This design aimed to detect the formation of hairpin loops by zipping of complementary base-pairs
at the ends. Such a process resulted in quenching of fluorescent intensity of $F$ by $Q$. They found 
that, for a given chain length, a polydA chain took longer time than a polydT chain to form a loop.
This indicates greater bending stiffness for polydA chains, resulting from stacking interaction between 
Adenine bases. More intriguing was the result that, loop closing time 
$\ln(\tau_c) \propto \beta$ for polydA and a nearly flat temperature dependence for polydT. 

This process can be approximately described by a two state system: a chain in a open state ($o$) 
or a chain with a closed loop($c$).
At equilibrium, the interconversion $o \rightleftharpoons c$ obeys a detail balance condition:
$\rho_c k_{c\rightarrow o}=\rho_o k_{o\rightarrow c}$ \cite{chen04}, where $\rho_{c/o}$ are the equilibrium
densities and $k_{c\rightarrow o},k_{o\rightarrow c}$ are the conversion rates. Loop closing
and opening times, $\tau_c$ and $\tau_o$, are inverses of the respective rates. Assuming a small
interaction radius $a$ between the chain  ends, approximately $\rho_c= \frac{4\pi}{3}a^3 P_N(\vec R=0)$,
where $P_N(\vec R)$ is the probability of finding the chain ends at a separation $\vec R$. But since 
possible number of open configurations far outweighs number of closed configurations i.e., $\rho_0\gg \rho_c$,
we can approximate $\rho_o=1-\rho_c\simeq 1$ \cite{chen04}. Further, $\tau_o$ is determined by the high energy 
barrier of the five bases long zipping strand, which is independent of the chain length $N$, and 
hence is a constant. Thus we arrive at $\tau_c\propto \rho_c^{-1}$, which we can compute from our 
model, as a function of chain length $(N)$ and temperature, at zero force. In Fig.2 we 
plot $\tau_c/N^{3/2}$ versus inverse temperature, obtained from our model and compare it with the
experimental data of Goddard et al \cite{goddard2000prl}. The rationale for rescaling $\tau_c$ by 
$N^{3/2}$ is to partially nullify the strong $N$ dependence in $\tau_c$. Although $\rho_c\propto 
N^{-3/2}$ only for an FJC model (at large $N$) and not for an WLC model, nevertheless it turns out 
to be useful in approximately collapsing both our simulation data and Goddard et al's experimental 
data in a narrow range of $N$, near the room temperature.
 
\begin{figure}[htb]
\includegraphics[width=8.5cm]{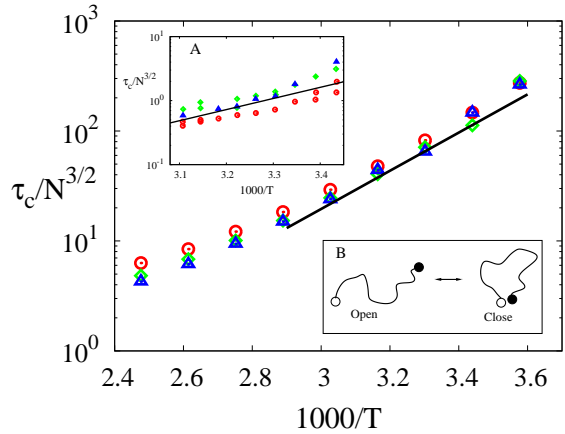}
\vspace{.25cm}
\caption{(color online) Semi-log plots of the scaled loop closing time $\tau_c$ (of polydA) versus inverse 
temperature, computed for various chain lengths $N=10$ (circle), $20$ (diamond) and $30$ (triangle), 
by Monte-Carlo simulation of our model. In a narrow window of temperature $1000/T \in [3.1,3.45]$, 
the scaling of $\tau_c$ has been experimentally shown (inset-A) to be approximately, Arrhenius type 
(i.e., $\tau_c \propto \exp(\epsilon/k_BT)$) by Goddard et al \cite{goddard2000prl}. Their data 
for chain lengths $N=12$ (circle), $21$ (diamond) and $30$ (triangle), are plotted after rescaling 
with $N^{3/2}$). Both the Arrhenius aspect
as well as the quantitative value of the exponent $\epsilon$ are closely reproduced by our model. 
The dashed lines in both the main plot and the inset have the same exponent 
$\epsilon=13.4k_BT_o$, where $T_o=298K$ is the room temperature. Inset-B schematically shows 
transition between open and closed chain conformations. 
}
\label{fig.AT}
\end{figure}

\vspace{.5cm}
\begin{figure}[htb]
\includegraphics[width=8.5cm]{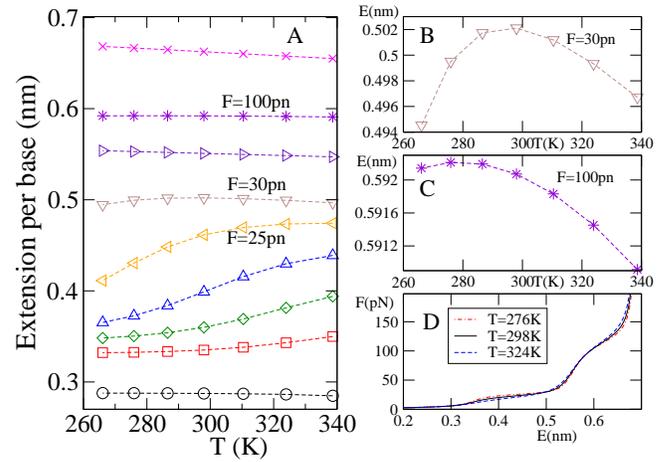}
\vspace{.25cm}
\caption{(color online) Extension(E) per base (in $nm$) versus temperature, at a fixed force, 
plotted in (A) for different 
values of forces: $5,10,15,20,25,30,60,100$ and $150pN$, from bottom to top.  These are obtained 
from our model using transfer matrix method. Effect of the first transition (near $25pN$)  
is reflected in the relatively large jump in the extension. Near the overstretching transition the
jump is not so pronounced due to the relatively small plateau width in the corresponding FX diagram.  
The variation of extension with temperature changes qualitatively with force. The extension weakly 
decreases with temperature at low force, increases with temperature at intermediate force ($<30pN$), 
shows interesting reentrant behavior at higher force 
($30-100pN$) and finally again decreases with temperature at very high force. The reentrant 
behavior i.e., extension initially increasing with temperature and later at higher temperature, 
decreasing with temperature, is zoomed in (B) and (C). The minute change in the FX curves with
temperature is shown in (D). 
Dashed lines are guide for the eyes.}
\label{fig.TvsF}
\end{figure}
{\em Temperature dependence of extension} :
Change of extension with temperature has been shown to give interesting behavior for wild type ssDNA
\cite{FvsT}. Fig.3 shows the analogous property resulting from our model of polydA, at fixed force, computed 
using transfer matrix method. In the absence of any transitions polymer extension is expected to 
decrease with temperature because of entropic elasticity, as it happens in rubber, for example. 
In case of wild type ssDNA hairpin loops can form which modifies the extension-temperature behavior 
in non-trivial ways.  For polydA although loops cannot form in the absence of complementary bases, 
existence of two transitions 
(helix-coil and overstretching) makes it behave in an interesting way (see Fig3A), to the extent 
that weak, nonmonotic, re-entrant behavior can be observed (Fig3B,C). These can be understood 
qualitatively. Two important points to remember here are, a) helix-coil transition can also be 
affected by raising temperature, and b) entropic fluctuations are enhanced at high temperatures 
which smoothens the force driven transitions (data not shown). Fig3A shows that at very low force 
like $5pN$, indeed 
extension weakly decreases with temperature but when force rises ($10pN$ onwards) as we move 
close to the helix-coil transition, it is easier to affect the transition by raising
temperature and helix-coil transition leads to rise in the extension.  But at forces just beyond 
the helix-coil transition (30pN onwards but much below 100pN) the ssDNA cannot access the 
overstretched bond lengths solely by means of thermal fluctuations and looses out to entropic 
elasticity showing decrease in extension with temperature. But as force approaches $100pN$ due
to its vicinity to the overstretching transition the extension again increases first but eventually
loose out to entropic elasticity at higher temperature, giving rise to reentrant behavior. 
Beyond the overstretching transition extension again decreases weakly with temperature due to 
entropic elasticity, weakly because at such high force not much entropy is left in the almost 
straight configuration.

In conclusion, we have proposed a new model for polydA, that incorporates two transitions 
and quantitatively reproduces both force-extension characteristics and loop closing statistics of such 
homopolymers. Our model also predicts interesting reentrant behavior in the temperature-extension 
diagram of polydA which can be verified experimentally.  
\begin{acknowledgments}
We thank Dibyendu Das for useful comments.
\end{acknowledgments}

\end{document}